\documentclass[reqno,12pt]{amsart}
\usepackage{amssymb,graphicx,url}
\usepackage{hyperref}
\begin{document}
\title{Autocorrelation type functions for big and dirty data series}
\author{Christoph Bandt}
\date{\today }
\maketitle

{\bf One form of big data are signals - time series of consecutive values.
In physical experiments, billions of values can now be measured within a second \cite{ar,sz}. Signals of heart and brain in intensive care, as well as seismic waves, are measured with 100 up to 1000 Hz over hours, days or even years \cite{physio,seis}. A song of 3 minutes on  CD comprises 16 million values. 

Music and seismic vibrations basically consist of harmonic oscillations for which classical tools like autocorrelation and spectrogram work well \cite{bd,shst,dr}.  This note presents similar tools for all kinds of rhythmic processes, with non-linear distortion, artefacts, and outliers.  Permutation entropy \cite{bp,zzrp,akk} has been used in physics \cite{sz,tk}, medicine \cite{sl,ng,fm}, and engineering \cite{nair,ylg}. Now ordinal patterns \cite{bs,lor,lrh} are studied in detail for big data.  As new version of permutation entropy, we define a distance to white noise consisting of four curious components.  Applications to a variety of medical and sensor data are discussed.}

\begin{figure}[h]
\includegraphics[width=.995\textwidth]{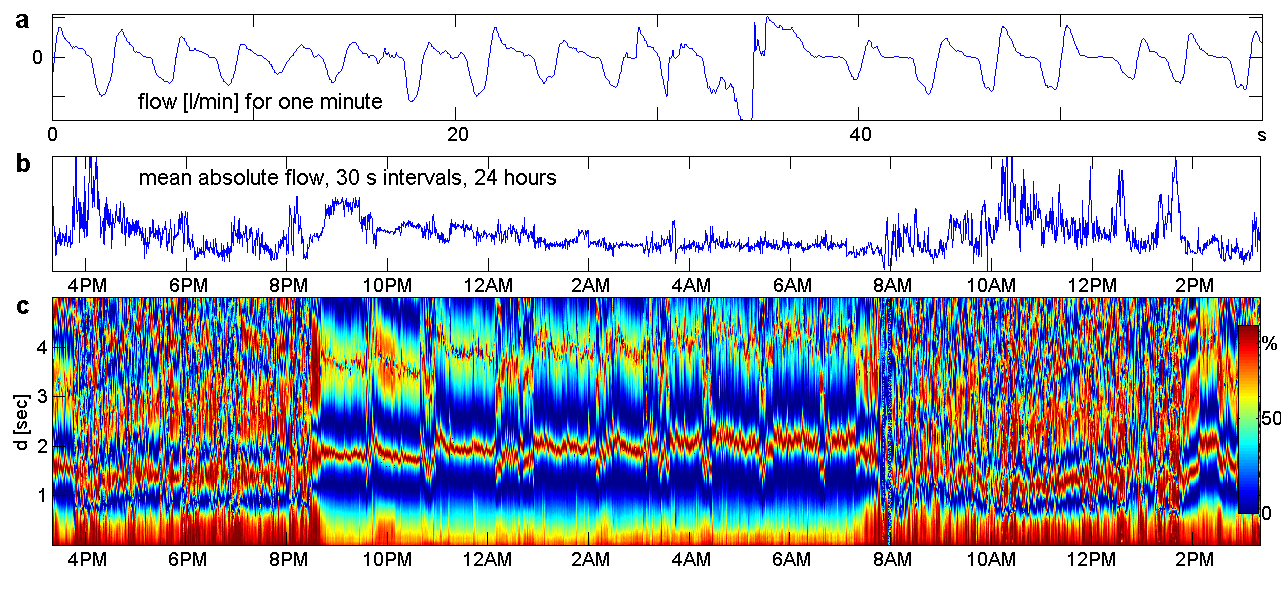}
\caption{Nose breathing of a healthy volunteer in normal life.   a: One minute of data without artefacts. b: Mean absolute flow for each 30 seconds of 24 hours measurement. c: New function $\tilde{\tau}(d)$ for each minute shows more structure, needs no calibration of data.}\label{nose}
\end{figure}

Classical time series from quarterly reports of companies, monthly unemployment figures, daily statistics of accidents etc. consist of 20 up to a few thousand values which were determined with great care.  Machine-generated data amount to millions and are not double-checked.  Usually it is easy to let the machine run faster and longer but processing goes the other way: data are smoothed out by downsampling.  This note shows how high time resolution can be
exploited with simple ordinal autocorrelation functions. To demonstrate the remarkable effect of these tools, they will be applied to raw data: no preprocessing, no filters, no detection of outliers, no removal of artefacts, and no polishing of the results.  

Figure 1 comes from ongoing work with Achim Beule (Department ENT, Head and Neck Surgery, University of Greifswald) on respiration of healthy volunteers in everyday life. Sensors measuring air flow with sampling frequency 50 Hz were fixed in both nostrils, gently enough to ensure comfort for 24 hours measurement. Mouth breathing was not controlled, so the signals contain lots of artefacts. Traditional analysis takes averages over 30 seconds to obtain a better signal. More information is found in a function $\tilde{\tau}(d)$ for each minute of the dirty signal. As explained below, $\tilde{\tau}$ measures the percentage of ''elliptical symmetry'' of the signal, depending on time and the delay $d$ which runs from 0.02 to 5 seconds. The collection of these functions,  visualized like a spectrogram, shows phases of activity and sleep, various interruptions of sleep, inaccurate measurements around 8 am, a little nap after 2 pm.  Frequency of respiration can be read from the lower dark red stripe which marks half of the wavelength, and the upper red and yellow stripe marks the full wavelength (4 seconds in sleep, less than 3 in daily life).

\subsection*{1. Up-down balance}
Only the order relation between the values of the time series $x=(x_1,...,x_T)$ will be used, not the values themselves. In the simplest case, the question is whether the time series goes more often upwards than downwards. For a delay $d$ between 1 and $T/2$, let $n_{12}(d)$ and $n_{21}(d)$ denote the number of time points $t$ for which $x_t<x_{t+d}$ and $x_t>x_{t+d},$ respectively.  We determine  the relative frequencies of increase and decrease over $d$ steps: $p_{12}(d) =n_{12}(d)/(n_{12}(d)+n_{21}(d))$ and
$p_{21}(d)=1-p_{12}(d).$  Ties are disregarded. The up-down balance 
\[ \beta(d)=p_{12}(d)-p_{21}(d) \quad \mbox{ for }\quad d=1,2,... \]
is a kind of autocorrelation function. It reflects the dependence structure of the underlying process and has nothing to do with the size of the data.

\begin{figure}[h]
\includegraphics[width=.99\textwidth]{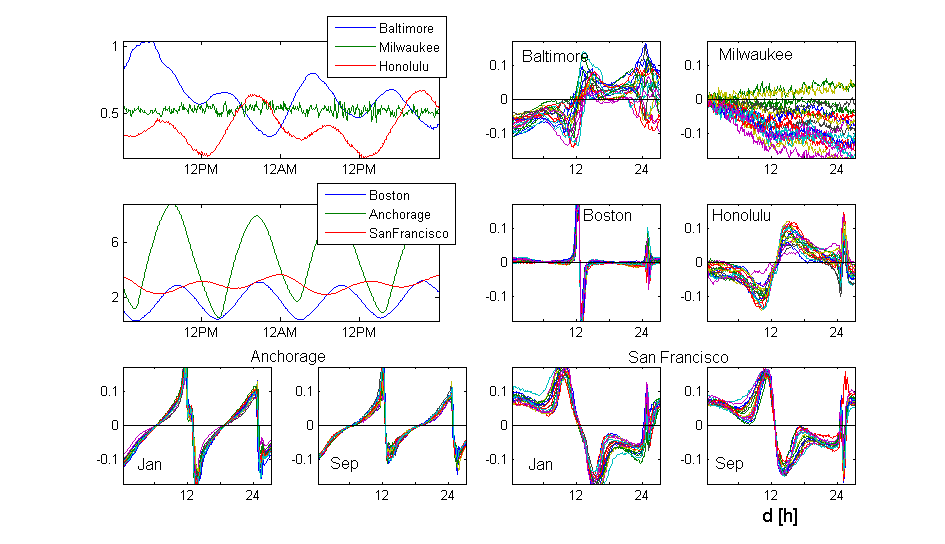}
\caption{Water levels at 6 minute intervals from \cite{water}. Original data shown for Sep 1-2, 2014. Functions $\beta$ are given for September of 18 years 1997-2014, and also for January in case of Anchorage and San Francisco. 
$d$ runs from 6 min to 27 hours.}\label{waterbeta}
\end{figure}

For illustration we take water levels from the database of the National Ocean Service \cite{water}.
Since water tends to come fast and disappear more slowly, we could expect $\beta (d)$ to be negative, at least for small $d.$  This is mostly true for water levels from lakes, like Milwaukee in Figure 2. For sea stations, tidal influence makes the data almost periodic, with a period slightly smaller than 25 hours.  The data have 6 minute intervals and we measure $d$ in hours, writing $d=250$ as $d=25 \rm h.$ A strictly periodic time series with period $L$ fulfils $\beta (L-d)=-\beta(d)$ which in the case $L=25$ implies $\beta(12.5)=0,$ visible at all sea stations in Figure 2. Otherwise there are big differences: at Honolulu and Baltimore the water level is more likely to fall within the next few hours, at San Francisco it is more likely to increase, and at Anchorage there is a change at 6 hours.  Each station has its specific $\beta$-profile, almost unchanged during 18 years, which characterizes its coastal shape. $\beta$ can also change with the season, but these differences are smaller than those between stations.

Figure 2 indicates that $\beta ,$ as well as related functions below, can solve some basic problems of statistics: describe, distinguish, and classify objects. Cf. Figure \ref{heyjude3}.

\subsection*{2. Sliding windows and statistical variation}
The time series is divided into pieces, so-called windows, and $\beta$ is determined for every window. If we have some hundred windows, which may overlap, we can compose them in a map like Figure 1 or 4, where each time point corresponds to a window, the $\beta$-functions are represented by vertical lines, and the values $\beta(d)$ are color coded.  Such maps show whether the regime of the underlying process changes in time, and indicate transitions between different regimes: activity and rest, or sleep stages.

If there are no systematic changes in the appearance of the functions, we can consider the underlying process as stationary and estimate statistical fluctuations. Mathematical arguments say that the standard deviation of $\beta(d)$ for fixed $d$ is  $\frac{c}{\sqrt{n}}$ where $n$ is the window length and $c$ some constant.  In our data, $c$ varied from $\frac12$ to 4, and in most cases we had $c\approx 2.$ 
 For the data of Figure 2, the underlying tidal process is certainly not stationary - but the influence of the moon was reduced by calculating $\beta$ over one month, the annual component was minimized by taking only September.  We can roughly estimate $\sigma ,$ with deviations still depending on $d.$  For medical data, errors were obtained from stationary segments, e.g. deep sleep. Another error of order $\frac{d}{n}$ is observed for very large parameters $d$ (Methods 2,3). 

The error estimate $\sigma\approx \frac{2}{\sqrt{n}}$ can be used to determine the appropriate window length. To have the $2\sigma$-radius of the $95\% $ confidence interval for $\beta(d)$ smaller than 0.01, we must take $n\ge 160000.$ For Figures 2 and 3 we have $n\approx 8000$ which gives a tolerance of $\pm 0.04.$ Thus
our method works \emph{only} for big data.

A simple exact method checks whether $\beta(d)$ significantly differs from zero, or the $\beta$-curves of two sites in Figure 2 significantly differ at some $d.$ When windows do not overlap,
the median test is applicable. Since $\beta(20)>0$ for all 11 pieces in Figure 3b, one concludes that the median of $\beta(20)$ for the underlying process cannot be zero since the $p$-value for 11 heads in 11 coin tosses is less than 0.001.
Thus, except for $\gamma ,$ most values of the functions in Figure 3 differ significantly from zero.  

\subsection*{3. Patterns of length 3}
Three equidistant values $x_t, x_{t+d}, x_{t+2d}$ without ties can realize six order patterns. 213 denotes the case  $x_{t+d}<x_t<x_{t+2d}.$
\[ \includegraphics[width=.65\textwidth]{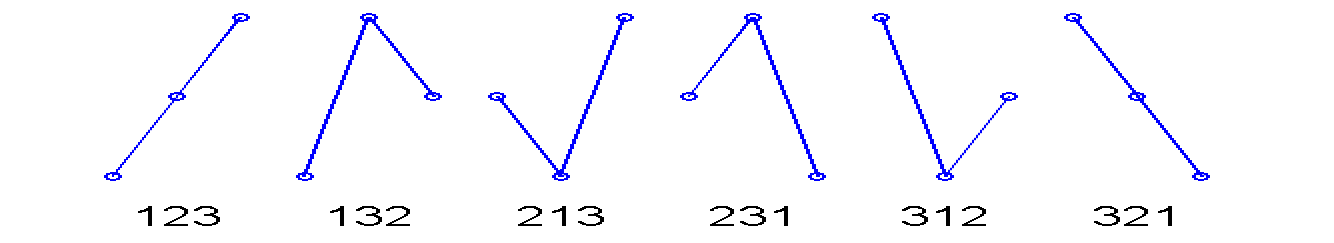} \]
For each pattern $\pi$ and $d=1,2,..,T/3$ we count the number $n_{\pi}(d)$ of appearances in the same way as $n_{12}(d).$  In case $\pi =312$ we count all $t=1,...,T-2d$ with $x_{t+d}<x_{t+2d}<x_t.$ Let $S$ be the sum of the six numbers. Patterns with ties are not counted.
Next, we compute the relative frequencies $p_\pi(d)=n_\pi(d)/S .$  For white noise it is known that all  $p_\pi(d)$ are $\frac{1}{6}$ \cite{bp}.

As autocorrelation type functions, we now define certain sums and differences of the $p_{\pi}.$ The function
\[ \tau(d)=p_{123}(d)+p_{321}(d)-\frac{1}{3} \]
is called persistence \cite{bs}.  This function indicates the probability that the sign of $x_{t+d}-x_t$ persists when we go $d$ time steps ahead. The largest possible value of $\tau(d)$ is $\frac{2}{3},$ assumed for monotone time series.  The minimal value is $-\frac{1}{3}.$ The constant $\frac{1}{3}$ was chosen so that white noise has persistence zero. The letter $\tau$ indicates that this is one way to transfer Kendall's tau to an autocorrelation function. Another version was studied in \cite{fgh}.

\begin{figure}[h]
\includegraphics[width=.47\textwidth]{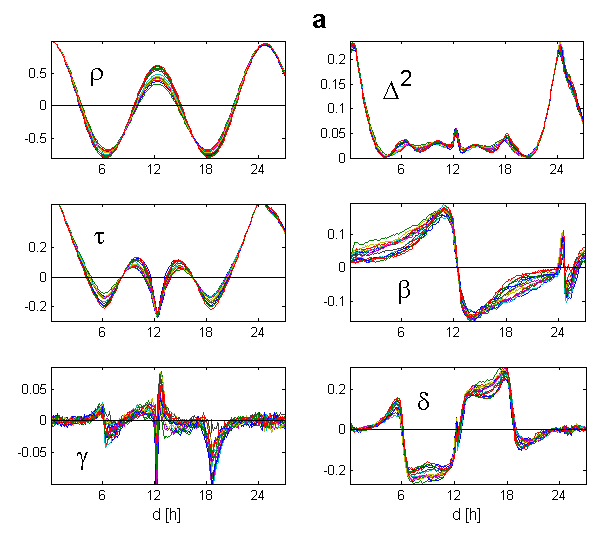}\quad  
\includegraphics[width=.47\textwidth]{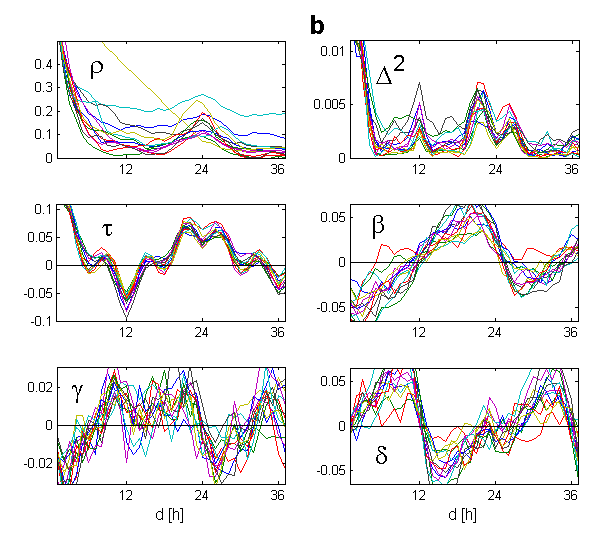}
\caption{Autocorrelation $\rho$ and ordinal functions $\Delta^2, \tau, \beta, \gamma, \delta$ for  {\bf a:} the almost periodic series of water levels at Los Angeles (September 1997-2014, data from \cite{water}), {\bf b:} the noisy series of hourly particulate values at nearby San Bernardino 2000-2011 from \cite{dust}, with weak daily rhythm, see Figure \ref{dustclust}.  The functions on the left are about three times larger, for $\Delta^2$ nine times, but fluctuations are of the same size. }\label{corrfcts}
\end{figure}

Similar to the classical autocorrelation function $\rho ,$ a period of length $L$ in a signal is indicated by minima of $\tau$ at $d=\frac{L}{2}, \frac{3L}{2}, \frac{5L}{2},...$  For these $d$ we have $x_t\approx x_{t+2d}$ so that the patterns 123 and 321 are rare.
Near to $d=L, 2L, 3L,...$ the function $\tau$ is large, as $\rho .$
For a noisy sequence, however, a local minimum appears directly at $d=L, 2L,...,$ since patterns tend to have equal probability there. The larger the noise, the deeper the bump. 
In Figure 3b, $\tau$ at $d=12$ and $d=24$ shows exactly this appearance and proves the existence of a 24 hour rhythm better than $\rho .$

Beside $\tau$ and $\beta$ we define two other autocorrelation type functions. For convenience, we drop the argument $d.$
\[ \gamma=p_{213}+p_{231}-p_{132}-p_{312} \]
is a measure of time irreversibility of the process, and
\[ \delta=p_{132}+p_{213}-p_{231}-p_{312} \]
describes up-down scaling since it approximately fulfils $\delta(d)=\beta(2d)-\beta(d)$ (Methods 3). 
Like $\beta ,$ these functions measure certain symmetry breaks in the distribution of the time series. They all have strong invariance properties (cf. Figure \ref{autopers}):

When a nonlinear monotonous transformation is applied to the data -- many sensors measure proxy effects which are monotonously, but not linearly related to the target variable -- all ordinal functions remain unchanged.
They are not influenced by low frequency components with wavelength much larger than $d$ which often appear as artefacts in data. 
Since $\tau, \beta, \gamma,$ and $\delta$ are defined by assertions like $X_{t+d}-X_t>0,$ they do not require full stationarity of the underlying process. Stationary increments suffice - Brownian motion for instance has $\tau (d)=\frac{1}{6}$ for all $d$ \cite{bs}.  Finally, ordinal functions are not influenced much by a few outliers. One wrong value, no matter how large, can change $n_\pi (d)$ only by $\pm 2$ while it can completely spoil autocorrelation. \vspace{2ex}

{\bf 4. Permutation entropy } is the Shannon entropy of the distribution of order patterns:
\[ H = -\sum_{\pi} p_{\pi} \log p_{\pi} \ .  \] 
$H$ can be defined for any level $n,$ using the vectors $(x_t,x_{t+d},...,x_{t+(n-1)d})$ and their $n!$ order patterns $\pi $ \cite{bp}.  In practice we hardly go beyond $n=7.$ Used as a measure of complexity and disorder, $H$ can be calculated for time series of less than thousand values since statistical inaccuracies of the $p_{\pi}$ are smoothed out by averaging. Applications include EEG data \cite{sl,ng,lor,fm}, optical experiments \cite{sz,ar,tk}, river flow data \cite{lrh}, control of rotating machines \cite{nair, ylg}, economic applications (cf. \cite{zzrp}), and the theory of dynamical systems \cite{am,uk}. Recent surveys on $H$ are \cite{zzrp,akk}. 
As a measure of disorder, $H$ assumes its maximum $\log n!$ for white noise. $D=\log n! -H$ is called divergence or Kullback-Leibler distance to the uniform distribution $p_\pi=\frac{1}{n!}$ of white noise.  

In this note all functions, including autocorrelation, measure the distance of the data from white noise. For this reason, we take divergence rather than entropy, and we replace
$-p_{\pi} \log p_{\pi}$ by $p_{\pi}^2.$ As before, we drop the argument $d.$ The function 
\[  \Delta^2 = \sum_{\pi} (p_{\pi} -{\textstyle \frac{1}{6}})^2 =  \sum_{\pi} p_{\pi}^2  -{\textstyle \frac{1}{6}} \]
where the sum runs over the six order patterns $\pi$ of length 3, will be called the \emph{distance of the data from white noise.} Of course, we can define  $\Delta^2$ for any length $n,$  replacing $\frac{1}{6}$ by $\frac{1}{n!} .$ More precisely, $\Delta^2$ is the squared Euclidean distance between the observed order pattern distribution and the order pattern distribution of white noise. Considering white noise as complete disorder, $\Delta^2$ measures the amount of rule and order in the data. The minimal value 0 is obtained for white noise, and the maximum $\frac{5}{6}$ for monotone time series.

From a practical viewpoint, $\Delta^2$ \emph{is just a rescaling of} $H,$ related to the quadratic Taylor approximation of $H$ around white noise parameters $p_\pi =\frac{1}{6}$  
\[ H \approx \log 6 - 3 \Delta^2 \ .\]
For our data these two functions can hardly be distinguished by eyesight.

\subsection*{5. Partition of the distance to white noise}
A Pythagoras type formula combines $\Delta^2$ with the ordinal functions: 
\[  4\Delta^2 = 3\tau^2 +2\beta^2 +\gamma^2 +\delta^2 \ . \]
This holds for each $d=1,2,...$ 
The equation is exact for random processes with stationary increments as well as for cyclic time series. The latter means that we calculate $p_{\pi} (d)$ from the series $(x_1,x_2,...,x_T,x_1,x_2,...,x_{2d})$ where $t$ runs from 1 to $T.$ For real data we go only to $T-2d$ and have a boundary effect which causes the equation to be only approximately fulfilled (Methods 4). The difference is smaller than 1\% in most of our data, see Figures \ref{no2Pleth},  \ref{heyjude2}, and \ref{tides}.  

This partition is related to orthogonal contrasts in the analysis of variance. When $\Delta^2(d)$ is significantly different from zero, we can define new functions of $d$:
\[ \tilde{\tau}=\frac{3\tau^2}{4\Delta^2}\ ,\quad  \tilde{\beta}=\frac{\beta^2}{2\Delta^2}\ ,\quad  \tilde{\gamma}=\frac{\gamma^2}{4\Delta^2}\ ,\quad  \tilde{\delta}=\frac{\delta^2}{4\Delta^2}\ . \]
By taking squares, we lose the sign of the values, but we gain a natural scale. $ \tilde{\tau},  \tilde{\beta},  \tilde{\gamma}$ and  $\tilde{\delta}$ lie between 0 and $1=100\% ,$ and they sum up to 1. For each $d,$ they describe the percentage of order in the data which is due to the corresponding difference of patterns.
Figure 1 shows $\tilde{\tau}$ for one-minute non-overlapping sliding windows as color code on vertical lines.

For Gaussian and elliptically symmetric processes, the functions $\beta, \gamma,$ and $\delta$ are all zero, and $\tilde{\tau}$ is 1, for every $d$ \cite[Section 5]{bs}. An image of $\tilde{\tau},$ as Figure 1, shows to which extent the data come from a Gaussian process and where the symmetry is broken. We get one percentage for each time and delay, 300000 values in Figure 1, and an overall average. As a rule of thumb, we exclude points for which $\Delta^2(d)<\frac{15}{n}$ where $n$ is window length (Methods 2). This bird's view method is a big data counterpart of rigorous tests for elliptical symmetry based on the covariance matrix \cite{hpa,ellip,omh}.  For ARMA processes, the average $\tilde{\tau}$ is above 97\%, as should be, for EEG data about 80\% , for music above 50\%, and for heart data only 30-50\%. The functions $\beta$ and $\delta,$ and less frequently $\gamma,$ can represent the main  part of $\Delta^2,$  but usually only for special values of $d.$ Details are discussed in the supplement.

\begin{figure}[h]
\includegraphics[width=.995\textwidth]{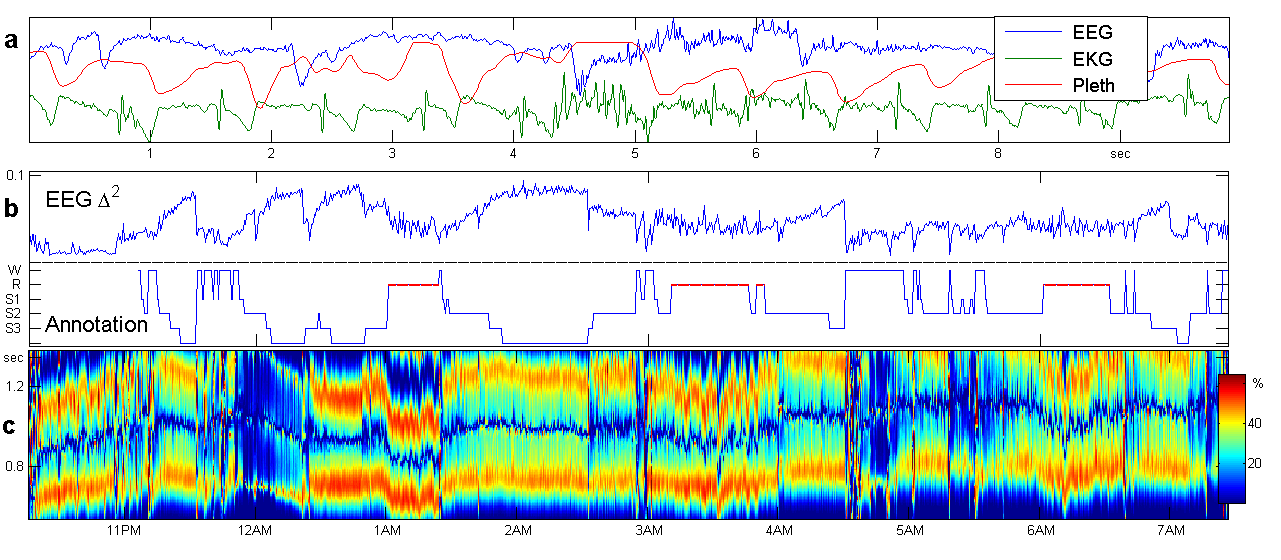} 
\caption{Sleep stages of healthy subject n3 from \cite{sleep}.   a: 10 seconds of data from EEG channel Fp2-F4, ECG and plethysmogram.  b: Expert annotation of sleep depth from \cite{sleep} agrees with distance $\Delta^2.$  c: Function $\tilde{\beta}$ of the plethysmogram describes  changes including REM.}\label{sleep}
\end{figure}

\subsection*{6. Sleep data} As an application, we study data from the CAP sleep database by Terzano et al. \cite{sleep} available at physionet \cite{physio}.
Sleep stages S1-S4 and R for REM sleep were already annotated by experts, mainly using the EEG channel Fp2-F4 and the oculogram.    Figure 4 demonstrates that permutation entropy $\Delta^2$ of that EEG channel, averaged over $d=2,...,20$ gives an \emph{almost identical estimate of sleep depth.} In \cite{ng,sleepscoring} permutation entropy was already recommended as indicator of sleep stages. We verified this coincidence in all data of the database with a normal EEG, which seems magic since $\Delta^2$ was introduced as a simple quantity, without any regard to sleep data.
For 512 Hz sampling frequency, $d=2,...,20$ corresponds to the time scale between 4 and 40 ms. Thus our $\Delta^2$ is a measure of smoothness of data at small scales which increases when high-frequency oscillations disappear.  This seems a complementary viewpoint to the classical R{\&}K rules for sleep stage classification and their recent modification \cite{sleepclassify} which refer to the appearance of low-frequency waves and patterns. 

REM phases are more difficult to detect. The oculogram was used for annotation. Figure 4c presents $\tilde{\beta}$
for the plethysmogram, measured with an optical sensor at the fingertip. We can see all interruptions of sleep and a lot of breakpoints which coincide with changes detected by the annotator. The REM phases are characterized by high values of $\tilde{\beta}(d),$ as well as an increase and a strongly increased variability of the heart rate:  the characteristic wavelength $d$ goes down. These variations influence ordinal functions so strongly that changes become more apparent than in a plot of heart rate.

Oximeters measuring the plethysmogram are cheap and easy to apply, in bed at home and possibly in daily life.  Many similar devices - armwrists, shirts, smartphone apps - are currently being developed.
Our ordinal functions can evaluate and visualize their high-resolution data. 

\subsection*{7. Conclusion}
On the basis of order patterns, error-resistent autocorrelation functions for big data were introduced. For patterns of length 3, the distance to white noise was divided into four interesting components. Various extensions seem possible. To develop a coherent theory of ordinal time series remains a challenge, despite groundbreaking work by Marc Hallin and co-authors \cite{hpu,fgh,hpa} on rank statistics. On the practical side, a spectrogram-like visualization was introduced which allows to see at one glance the course of heart and respiration data over 24 hours (Figures 1,4). Simple as it is, this technique may apply to data of sensors on satellites, weather stations, factory chimneys etc. as well as to experiments on nanoscale \cite{ar,tk} which may lay the ground for future generations of sensors. In any application, the methods presented here have to be modified, combined with established techniques, specific machine-learning and imaging tricks. 
Mathematical extraction of information must keep track with the revolution in sensor and computer technology: this seems an emerging field of study.

\pagebreak

\section*{Methods}
\subsection*{1. The program code}
A few lines of MATLAB code provide all our functions. {\tt x(1:T)} denotes the given time series, {\tt d} is between 1 and dmax.
\\{\tt y0=x(1:T-2*d); y1=x(d+1:T-d); y2=x(2*d+1:T);\\
s=2*((y0>y1)+(y0>y2))+(y1>y2); }\\
{\tt s} is a vector which contains the symbols $0,1,...,5$ representing the six order patterns of $(x_t,x_{t+d},x_{t+2d})$ for $d=1,...,T-2d.$ We correct for missing values NaN and ties by assigning symbols 6 to 11:
\\ {\tt h=isnan(y0)|isnan(y1)|isnan(y2)|(y0==y1)|(y0==y2)|(y1==y2);\\  s=s+6*h;}\\
Here $|$ means 'or'. If patterns with equality are to be counted, this line has to be modified. We now use the histogram function to determine the frequencies of the six patterns.
\\ {\tt p=hist(s,12)/(T-2*d-sum(h)); q=p(1:6); }\\
This will determine all our functions, for example:
\\ {\tt beta=p(1)-p(6); tau=p(1)+p(6)-1/3; 
h2=(q-1/6)*(q-1/6)'; H=-log(q)*q';}\\
On a PC, the algorithm obtains Figure 3 within a few seconds and Figure 1 within a few minutes.  In some papers \cite{akk}, permutation entropy of scale $d$ uses patterns for sums of $d$ consecutive terms of the $x_t,$ which makes sense if the $x_t$ represent a density function, like precipitation or workload on a server. This version is easily implemented by   cumulative sums, adding {\tt x=cumsum(x)} as first line to the program.

\subsection*{2. Statistical accuracy}
We consider a fixed $d$ and windows of length $n.$ Then $p_{12}(d)$ is a random sum of $n$ terms 1,0 which say whether $x_t<x_{t+d}$ is true or false. According to the binomial model, the standard deviation of $p_{12}(d)$ and $\beta(d)= 2p_{12}(d)-1$ would be $\frac12\frac{1}{\sqrt{n}}$ and $\frac{1}{\sqrt{n}},$ respectively. In reality, the variation is bigger because the terms in the sum are correlated. The correlations between differences $Y_t=X_{t+d}-X_t$ are relevant. As a simple model, we take the sum of $\frac{n}{k}$ independent terms 1,0 each of which is repeated $k$ times. For $\beta(d)$ this gives the standard deviation $\frac{c}{\sqrt{n}}$ with $c=k\sqrt{k}.$ An estimate of $c$ can only come from data. As a rule we got $c\approx 2.$ For $\tau$ the variation is a bit smaller, for $\gamma$ and $\delta$ a bit larger. Note that correlations can increase with $d.$ The partition formula and some assumptions give a rough estimate of $\frac{7}{n}$ for the standard deviation of $\Delta^2.$ Twice the value, $\frac{15}{n},$ is taken as a kind of confidence bound for the hypothesis $\Delta^2=0.$ This is just a guideline: for the data in the supplement it works well.

\subsection*{3. Identities for pattern frequencies and boundary effect} We fix $d$ and consider the $p_\pi$ for the whole time series. The equation  $p_{12}= p_{123}+p_{231}+p_{132}$ holds since on the right we count all $t=1,...,T-2d$ with $x_t<x_{t+d}.$ Similarly,   $p_{12}= p_{123}+p_{213}+p_{312}$ because on the right we count all $t=d+1,...,T-d$ with
$x_t<x_{t+d}.$ Strictly speaking, both relations are not correct since the count for $p_{12}$ is over $t=1,...,T-d.$ The largest possible difference is $\frac{d}{T-d}$ when we exclude ties for simplicity. This upper bound is negligible for large $T$ and comparably small $d,$ and random fluctuations usually make the difference much smaller than the upper bound.
Taking the difference of the two equations we see that the quantity
\[ \epsilon = p_{231}+p_{132}-p_{213}-p_{312} \]
equals zero, with the same degree of accuracy. In other words, the numbers of local maxima and of local minima in a time series coincide. Adding both identities for $p_{12}$ and subtracting $1=\sum p_\pi$ we obtain
\[ \beta =p_{123}-p_{321} \]
which has been used throughout the paper to calculate $\beta ,$ see the code above. Another relation of this type is $p_{12}(2d)= p_{123}(d)+p_{132}(d)+p_{213}(d).$ As a consequence, we get
 $\beta(2d)=2p_{12}(2d)-1=\beta(d)+\delta(d) $
which says that the functions $\beta$ and $\delta$ are tightly connected.
 For the mathematical model of processes with stationary increments, as well as for the cyclic time series mentioned in section 5, all these equations are exact \cite{bs}. 

\subsection*{4. Proof of the $\Delta^2$ partition formula} 
As in the code, we set $q_1=p_{123}-\frac{1}{6}, q_2=p_{132}-\frac{1}{6}, q_3=p_{213}-\frac{1}{6}, q_4=p_{231}-\frac{1}{6}, q_5=p_{312}-\frac{1}{6}, q_6=p_{321}-\frac{1}{6}.$ By high-school algebra
\begin{eqnarray*} 4 \sum q_k^2=&2(q_1+q_6)^2+2(q_1-q_6)^2&+(q_2+q_3+q_4+q_5)^2+(q_2-q_3-q_4+q_5)^2\\ &&+(q_2+q_3-q_4-q_5)^2+(q_2-q_3+q_4-q_5)^2 \end{eqnarray*}
Since $\sum q_k=0,$ the third square on the right is the same as the first. Using the definitions of $\Delta^2, \tau, \gamma, \delta$ and the identities above for $\beta$ and $\epsilon$  we get
\[ 4\Delta^2= 3\tau^2 +2\beta^2+\delta^2+\gamma^2+\epsilon^2 \]
which implies the formula because $\epsilon=0.$ The equation is exact for cyclic time series, and for random processes with stationary increments. For ordinary time series there is a small error, as discussed above, and we must check the accuracy of the equation from the data, see Figures \ref{no2Pleth},  \ref{heyjude2}, and \ref{tides} where differences are below 1\% .

\subsection*{5. Contents of supplement}
We give numerical and visual evidence for our partition formula and check our bound for small $\Delta^2.$ We explore the potential of our methods for various applications - medical data, speech, weather and ecological data - and test how far we can decrease the window size.  Autocorrelation and persistence are compared in Figure \ref{autopers} for a simple AR2-process, in Figure \ref{heyjude} for a song of The Beatles, and in Figure \ref{laser3500} for a laser experiment of Sorriano et al.~\cite{sz}. Further details of ordinal autocorrelation functions are discussed. 

\subsection*{Acknowledgement.} I am grateful to Bernd Pompe, Marcus Vollmer, Luciano Zunino, Mathias Bandt, Petra Gummelt and Helena Pe\~na  for suggestions and comments on drafts of this paper.\vspace{3ex}\\
{Institute of Mathematics\\ University of Greifswald\\ 17487 Greifswald, Germany\\}
\url{bandt@uni-greifswald.de}\pagebreak

\section*{Supplement}
\begin{figure}[h]
\includegraphics[width=.99\textwidth]{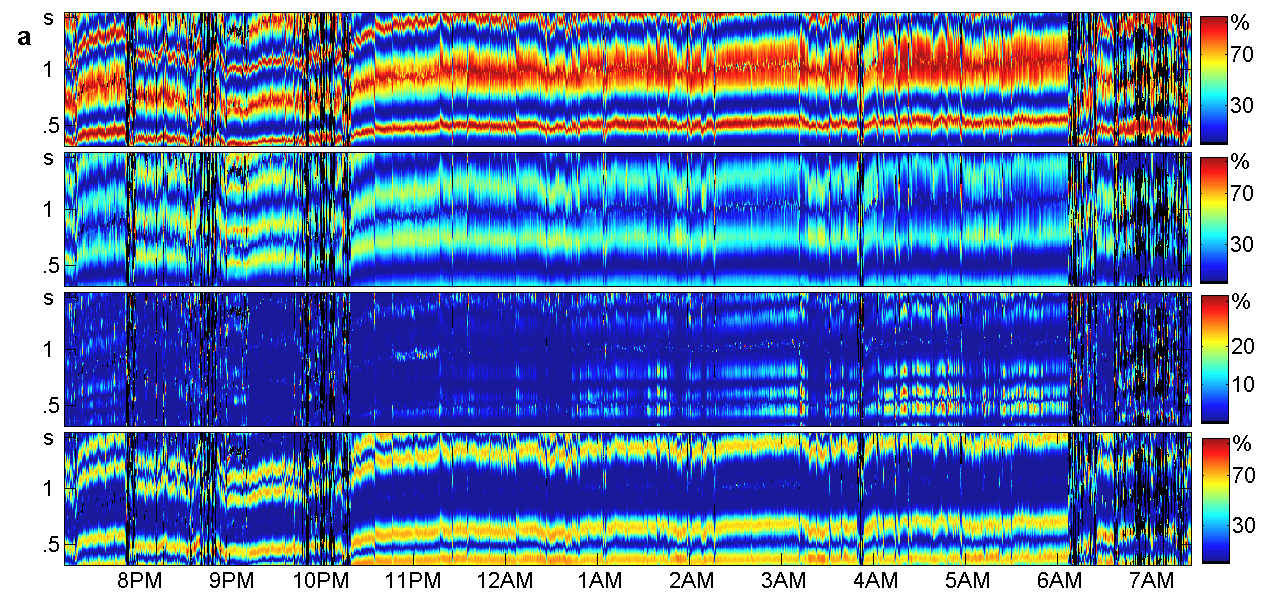}
\includegraphics[width=.99\textwidth]{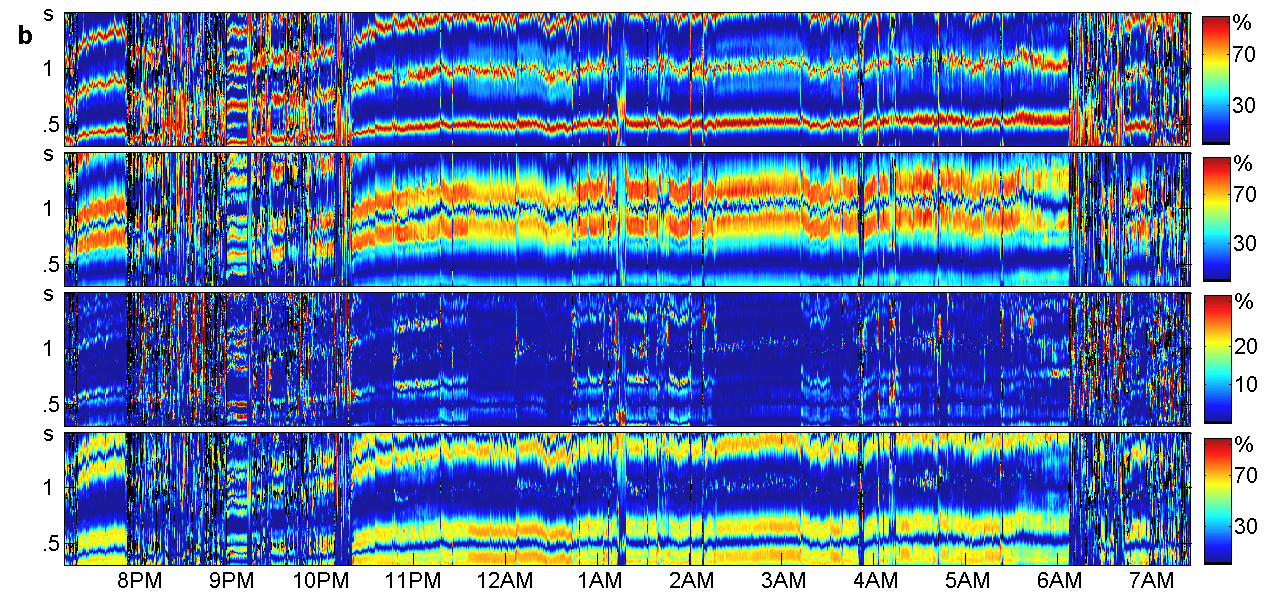}
\caption{Components of $\Delta^2$ for {\bf a} the plethysmogram and {\bf b} the ECG of control n2 of \cite{sleep}, with $d$ from 0.3 to 1.5s.   {\bf a} from above: $\tilde{\tau}$ with overall percentage 46\% ,  $\tilde{\beta}$ with 25\% , $\tilde{\gamma}$ with 3\% (colors scaled by factor 3), and  $\tilde{\delta}$ with 26\%  . The remainder is only $0.07\% $ in average. Places with $\Delta^2<\frac{15}{n}$ were not included in the average. They account for 7\% of all places and are marked black in the picture. The vast majority of black spots is caused by artefacts, movements of the patient.
In {\bf b} we have $\tilde{\tau}$ with average 31\%,  $\tilde{\beta}$ with 35\%, $\tilde{\gamma}$ with 5\% (colors scaled by factor 3),   $\tilde{\delta}$ with 29\%,  and $0.34\% $ remaining error. The percentage of black spots was 5\% . Can you guess REM phases?  Solution in Figure \ref{entrosleep}.} 
\label{no2Pleth}
\end{figure}

\begin{figure}[ht]
\includegraphics[width=.99\textwidth]{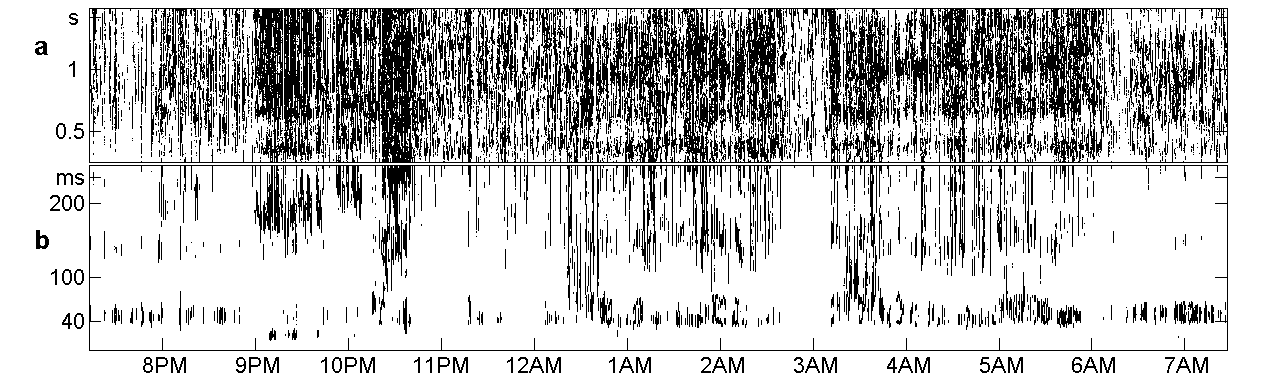}
\caption{Places with $\Delta^2<\frac{15}{n}\approx 0.001$ in the EEG record of n2 are marked black.  {\bf a.} For $d=0.25...1.5s$  no structure can be seen. {\bf b.}  For $d\le 0.25s$ the light places are related to stages of deep sleep in the figure below.  }  \label{eegtau}\vspace{3ex}
\includegraphics[width=.99\textwidth]{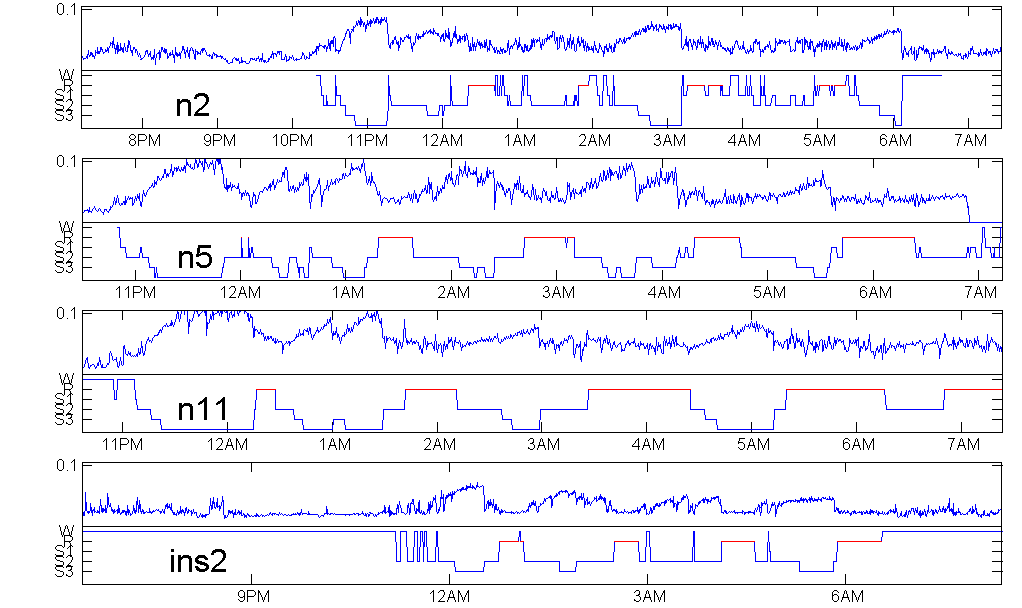}
\caption{Coincidence of mean $\Delta^2$ of EEG and annotation of sleep stages, for three controls from \cite{sleep} and one patient suffering from insomnia, with smaller $\Delta^2.$ Data were selected by record structure, not by the coincidence!
There are commercial systems for automatic sleep scoring. But $\Delta^2$ is such a simple, transparent quantity that this coincidence is worthwhile to study. Mean was taken over $d=2,...,20,$ that is 4 to 40ms  for 512Hz records.  
  }  \label{entrosleep}\vspace{10ex}
\end{figure}

\begin{minipage}{\textwidth}
\centerline{\bf 1. Overview\vspace{1ex}}
We start with a list of our datasets.
The essential parameters are window length and mean $\Delta^2$ which quantifies the amount of structure in the data. The latter is not so easy to fix since it depends very much on $d.$  For $d$ near zero, autocorrelation, persistence and hence also $\Delta^2$ assume very large values.\vspace{1ex}

\begin{center}\begin{tabular}{|l|c|c|c|c|}\hline 
subject / data &window& range of $d$ & mean $\Delta^2$ & motivation / application\\ \hline\hline
plethysmogram&3840/ 30s&0.3-1.5s&0.06&proof of partition formula \\ 
ECG&15360/30s&0.3-1.5s&0.04&visualize long-term data\\ \hline
EEG&15360/30s&0.25-1.5s&0.002&screening for structure\\ 
& &0-0.25s&0.013&formula for sleep scoring\\ \hline
AR2 process&2000&1-100&0.01&theoretical model\\ \hline
music/speech&2205/50ms&0-7ms&0.024&speech analysis/synthesis\\ \hline
temperature&720/1month&1-50h&0.03&exploratory data analysis \\ \hline
particulates&1200/50days&1-72h&0.003&notoriously noisy data\\ \hline
tides&1242/5days&10.5-14.5h&0.10&visualization of dynamics\\ \hline
laser data&3520/88ns&741-800&0.01&detection of period\\ \hline
\end{tabular}\vspace{2ex}\\ 
Table 1. Data studied in this supplement 
\vspace{3ex}\end{center}

\centerline{\bf 2. Heart and brain data\vspace{1ex}}
Figure \ref{no2Pleth} verifies the partition formula for heart data, and approves our 'confidence bound'  $\frac{15}{n}$ for  $\Delta^2=0.$ The vast majority of excluded places comes from artefacts, and it would make no sense to exclude still more places since the partition equation holds true with such a small error.  We note that in millions of places in Figure \ref{no2Pleth} and similar data,  $\tilde{\tau}+ \tilde{\beta}+ \tilde{\gamma}+\tilde{\delta}$ was always
smaller than 100\% plus a rounding error of $10^{-10}\,\% .$ This inequality seems to be generally true.\vspace{2ex}

Processes with small average $\Delta^2$ are more difficult to handle. Figure \ref{eegtau} concerns the EEG channel Fp2-F4 of the same record, n2 of \cite{sleep}, sampled with 512Hz. For $d=0.25,...,1.5s$ the average $\Delta^2$ is only 0.0017, and 47\% of the places have small $\Delta^2.$ They are spread rather uniformly, so there is little chance to get information from this range of $d.$ We concentrate on $d\le 0.25s$ with average $\Delta^2$ about 0.013 and only 10\% places below the bound. There seems some structure connected with sleep annotations. We have chosen the white field near the line $d=0$ when we average $\Delta^2$ over $d=4,...,40$ms to classify sleep stages. As demonstrated in Figure \ref{entrosleep}, this rough formula works perfectly well. But Figure \ref{eegtau} indicates other options for a better formula, which certainly can be found by looking more carefully at many data.
\end{minipage}

\vspace{2ex}

\begin{figure}[ht]
{\bf 3. A Gaussian model process}\vspace{1ex}

\begin{center}
\includegraphics[width=.69\textwidth]{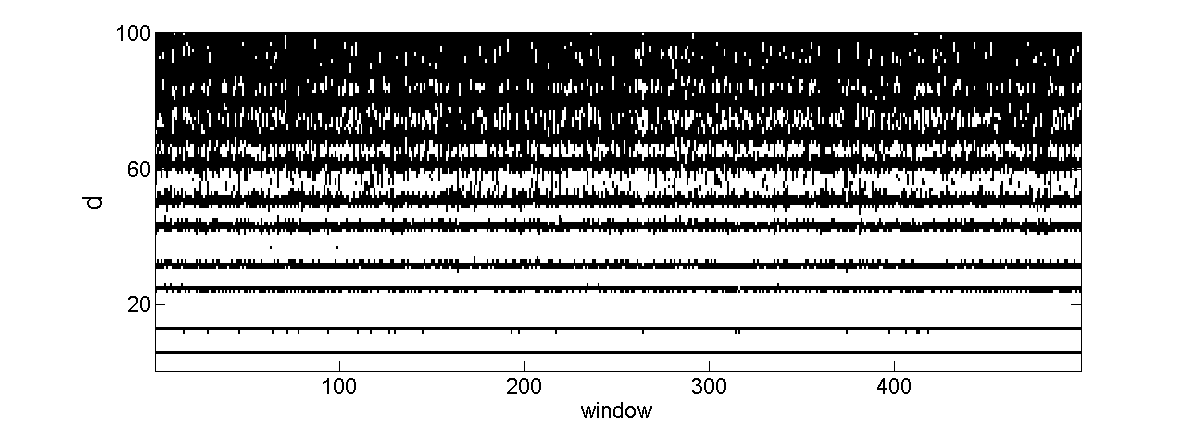}
\caption{Places with $\Delta^2<0.0015$ for an AR2 process. For large $d$ they are everywhere, for small $d$ they become concentrated on lines where $\tau (d)\approx 0.$ }  \label{ar2smallplaces}\vspace{2ex}
\end{center}

\begin{minipage}{0.95\textwidth}
AR2 processes belong to the simplest stationary Gaussian processes. Autocorrelation and persistence can be determined analytically \cite{shst,bs}. Here we  use the same procedures as for our data, to find out whether $\tilde{\tau}$ will be really 100\%. We take the process $X_t=1.85 X_{t-1}-0.96 X_{t-2}+W_t$ with Gaussian noise $W_t$ which has oscillating and not too rapidly decreasing autocorrelation (Figure \ref{autopers}). Dependencies for $d>100$ are very small, so we consider $d$ from 1 to 50 or 100. Figure  \ref{ar2smallplaces} shows that small places for $\Delta^2$ form horizontal stripes where $\tau (d)\approx 0.$
We determine $\tilde{\tau}$ on all places and on large places 
for several thousands of samples of two sizes $n.$
\vspace{2ex}

\begin{center}\begin{tabular}{|l r|c|c|c|}\hline 
window & delays&$\Delta^2<\frac{15}{n}$&$\tilde{\tau},$ all&$\tilde{\tau},$ corrected\\ \hline
$n=10000$& $d\le 100$& 49\% & 86\% & 98.4\% \\ 
&$d\le 50$ \ & 20\% & 95\% & 98.9\% \\ \hline
$n=2000$& $d\le 100$& 76\% & 73\% & 97.7\% \\ 
&$d\le 50$ \ & 55\% & 87\% & 97.9\% \\ \hline
\end{tabular}\vspace{0.5ex}\\ 
Table 2. The partition formula is not useful for processes with small $\Delta^2.$\vspace{2ex}\end{center}

Increasing our bound $\frac{15}{n}$ would only slightly improve $\tilde{\tau},$ on the cost of excluding many other places. Thus our partition formula seems not so useful for small average $\Delta^2,$ in this case about $0.01.$ Here $\tau$ is better to work with than $\tilde{\tau}.$ \vspace{2ex}

{\bf Remark.} At this point, let us briefly explain the type of symmetry given when ordinal functions are zero. We take a random process $(X_t)$ with stationary increments, fix a delay $d$ and consider the two-dimensional distribution of $Z_1=X_{t+d}- X_t$ and $Z_2=X_{t+2d}- X_{t+d}.$ The frequencies of order patterns are probabilities of sectors in the $z_1z_2$-plane: $\pi_{123}(d)=P(Z_1>0,Z_2>0), \pi_{321}(d)= P(Z_1<0,Z_2<0)$ and $\pi_{132}(d)=P(Z_1+Z_2>0,Z_2<0), \pi_{213}(d)=P(Z_1<0,Z_1+Z_2>0), \pi_{231}(d)=P(Z_1>0,Z_1+Z_2<0), \pi_{312}(d)=P(Z_1+Z_2<0,Z_2>0).$
When $(Z_1,Z_2)$ has a density function $f$ which is symmetric to the origin, i.e. $f(z_1,z_2)=f(-z_1,-z_2)$ then $\beta(d)=\delta(d)=0.$  If $f$ is symmetric to the line $z_1+z_2=0,$ i.e. $f(z_1,z_2)=f(-z_2,-z_1)$ then $\gamma(d)=\beta(d)=\delta(d)=0.$ Ordinal functions measure deviations from this kind of symmetry. 
\end{minipage}
\end{figure}

\begin{figure}[ht]
\begin{minipage}{0.95\textwidth}
In the figure below, we study how autocorrelation and persistence of the above AR2 process will change when we add various disturbances to the data. While autocorrelation works better when data are perturbed by white noise, persistence performs better for other perturbations which occur in practice. In particular, data from a sensor with nonlinear characteristics need not be calibrated when we use ordinal functions. We note that for cases {\bf c, d,} and {\bf e} in Figure \ref{autopers}, values of $\tilde{\tau}$ from Table 2 remain above 97\% .
\vspace{2ex}\end{minipage}

\includegraphics[width=.99\textwidth]{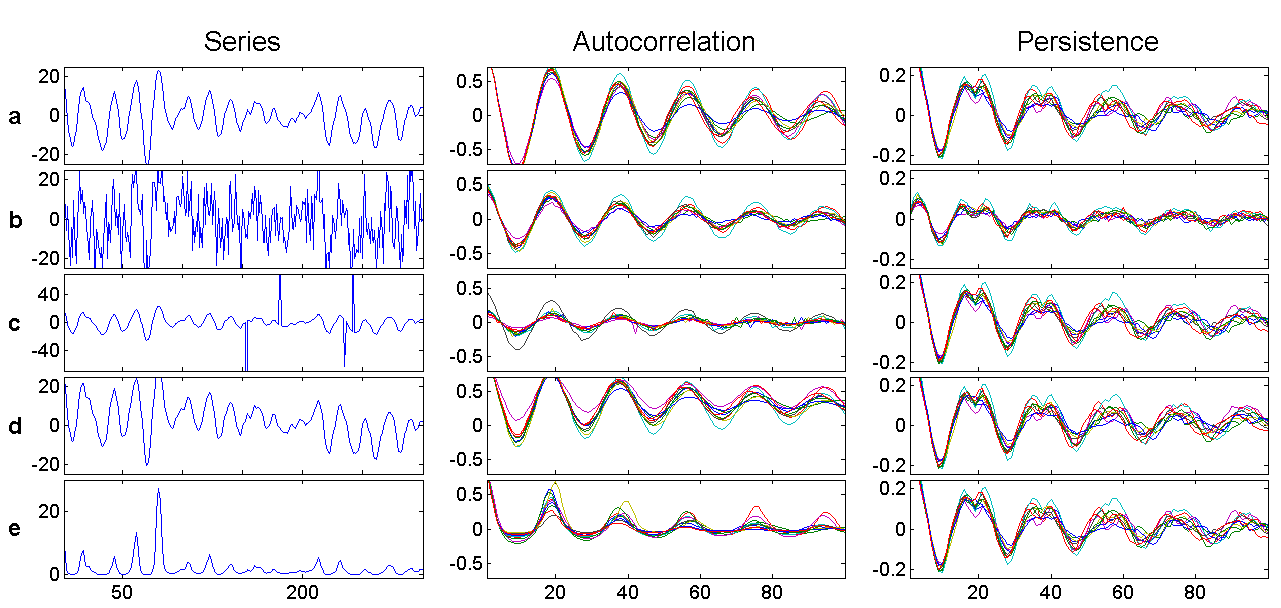}
\caption{Autocorrelation and persistence for an AR2 process with additive disturbances.  In each case 10 samples of 2000 points were processed, to determine statistical variation. On the left, 300 points of the respective time series are sketched. {\bf a.} Original signal.  {\bf b.} Additive Gaussian white noise with signal-to-noise ratio 1 divides autocorrelation by 2 and makes persistence very flat.  {\bf c.} 1\% outliers with average amplitude 20 times $\sigma$ of the original signal will harm the autocorrelation but not persistence. {\bf d.} Low-frequency function $\sin t/300$ added: autocorrelation becomes mostly positive, persistence keeps its shape. {\bf e.} The monotonous transformation $y=e^{x/7}$ is applied to the values: autocorrelation is distorted, persistence unchanged. }  \label{autopers}\vspace{10ex}
\end{figure}

\begin{figure}[ht]
{\bf 4. Speech and music}\vspace{1ex}
\begin{minipage}{0.95\textwidth}
For music data, the autocorrelogram sometimes shows the melody better than the spectrogram. Persistence behaves similarly.
\vspace{2ex}\end{minipage}

\includegraphics[width=.99\textwidth]{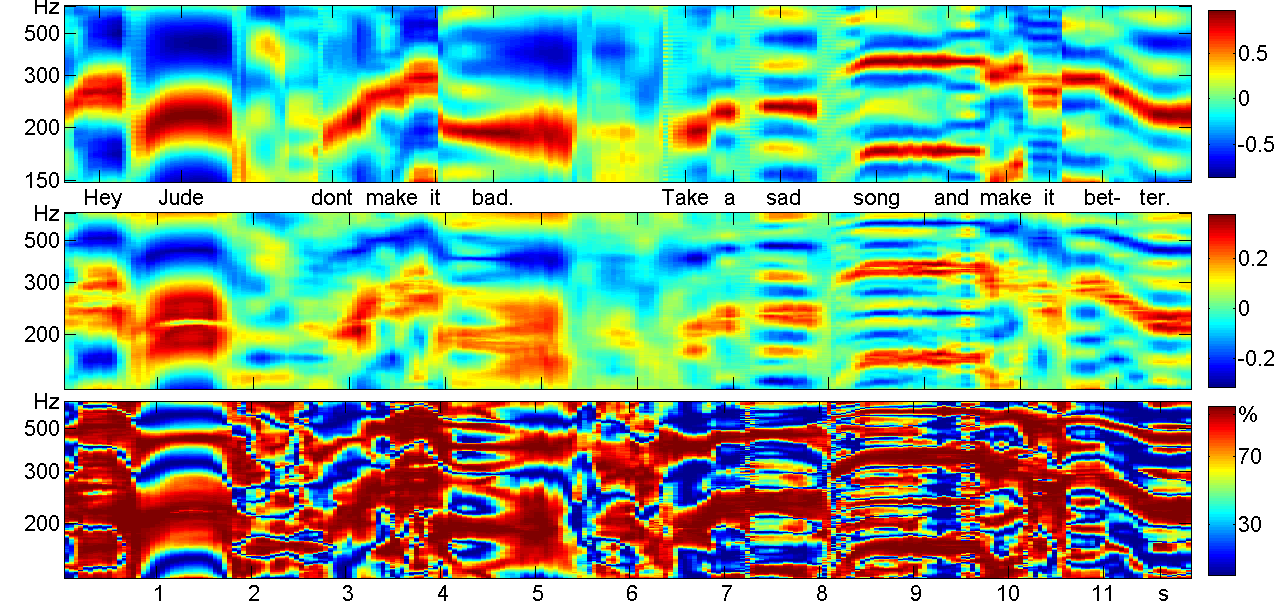}
\caption{The first 12 seconds of the song ''Hey Jude'' of The Beatles in sliding window analysis. Autocorrelation and persistence were calculated for overlapping windows of 0.25 seconds length, $n=11025.$ \ Reversed $y$-axis  helps recognize the melody, $d=49,...,294$ corresponds to 900,...,150\, Hz. Persistence gives a very similar picture as autocorrelation, except for the local minima at full periods. The function $\tilde{\tau}$ at the bottom has quite a number of blue spots where $\tilde{\beta}$ and $\tilde{\delta}$ yield the main contribution to $\Delta^2 .$
}  \label{heyjude}\vspace{3ex}

\begin{minipage}{\textwidth}
If we like to recognize not only melody, but also the text, we have to work with short windows, as shown in Figure \ref{heyjude2}. The ordinal structure of speech seems an interesting object for its own sake, as well as for applications.\vspace{1ex} 

In Figure \ref{heyjude4}, places with small $\Delta^2$ indicate unvoiced sounds when they are arranged in a vertical pattern. When they form horizontal patterns, which are somewhat wavy in case of a song, they indicate lines $\tau (d)\approx 0$ within voiced sounds. Otherwise they just indicate noise. Voiced sounds can be identified best by their prominent $\beta, \delta,$ and $\gamma$-shapes, as shown in  Figure \ref{heyjude3}.
\end{minipage}
\end{figure}

\begin{figure}[h]
\includegraphics[width=.99\textwidth]{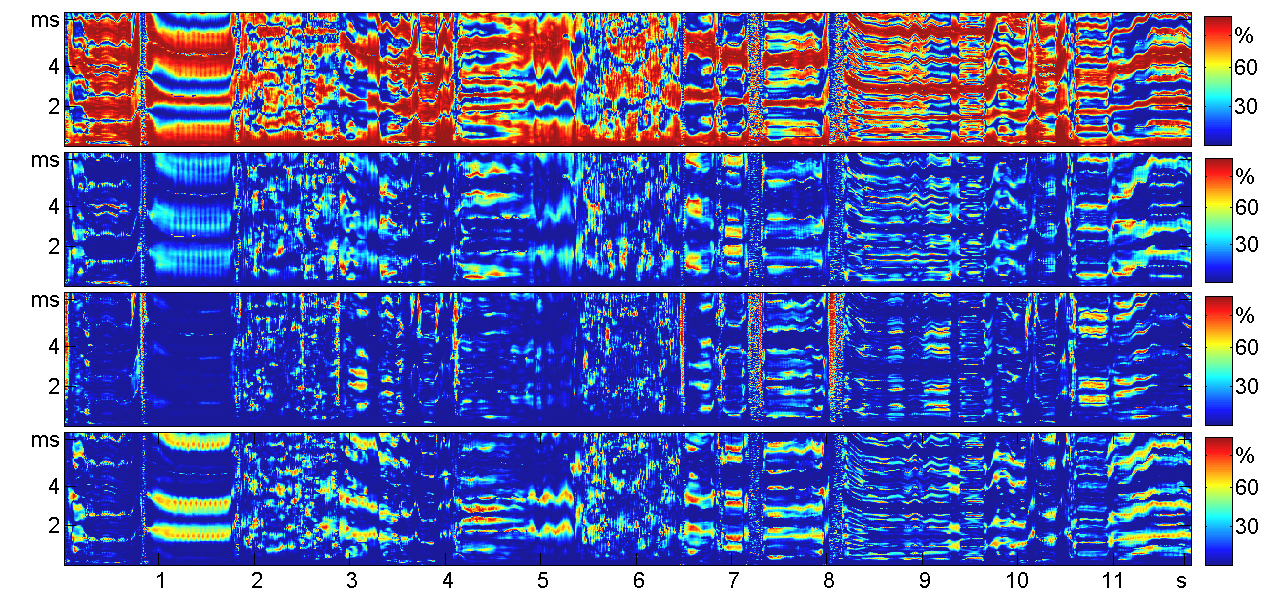}
\caption{The first 12 seconds of ''Hey Jude''  as partition of $\Delta^2$ with $\tilde{\tau}, \tilde{\beta}, \tilde{\gamma},$  and $\tilde{\delta}$ (from above).  A window of length 50ms was shifted in steps of 10ms. The $y$-axis is in normal orientation, $d$ runs from 1 to 294, that is, 0.02 to 6.7ms. The short window  reveals a lot of detail structure. All vowels show clear $\beta$ and $\delta$-patterns, even $\gamma$ is relevant at several places. The overall percentage is 57\%  for $\tilde{\tau},$  15\%  for $\tilde{\beta},$ 11\%  for  $\tilde{\gamma,}$ and 16\%  for $\tilde{\delta}.$ The mean of the rest, i.e. the error of the partition formula, is 0.77\% which is good for the short window. Places with small $\Delta^2$ excluded from this calculation are discussed below. \vspace{4ex} } \label{heyjude2} 
%
\includegraphics[width=.96\textwidth]{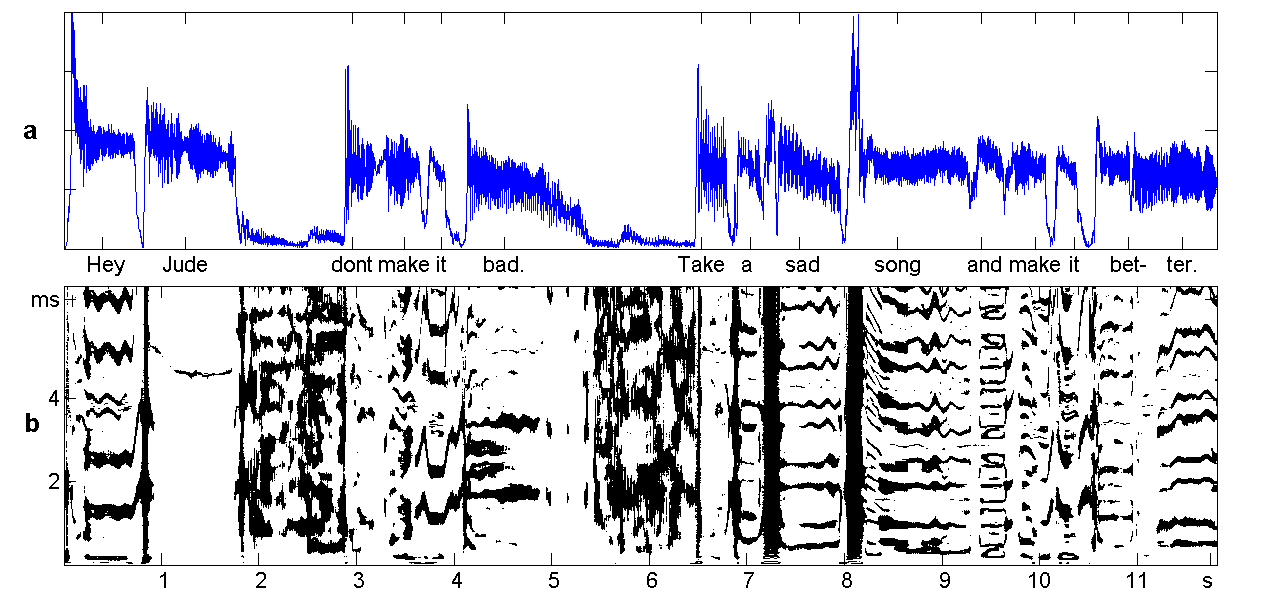}
\caption{In the previous figure, 26\% of the places $(t,d)$  fulfil $\Delta^2(d)<\frac{15}{n}\approx 0.007$ so that the signal does not significantly differ from white noise.  They form the dark spots in {\bf b.} For comparison, {\bf a} shows the mean absolute amplitude of the signal. We have small $\Delta^2$ for the two breaks with small amplitude, but much more pronounced for the unvoiced sounds, notably 's' in 'sad' and 'song'. In the voiced sounds, places with small $\Delta^2$ appear for $d$ with persistence zero. 
} \vspace{2ex} \label{heyjude4} 
\end{figure}

\begin{figure}[h]
\includegraphics[width=.99\textwidth]{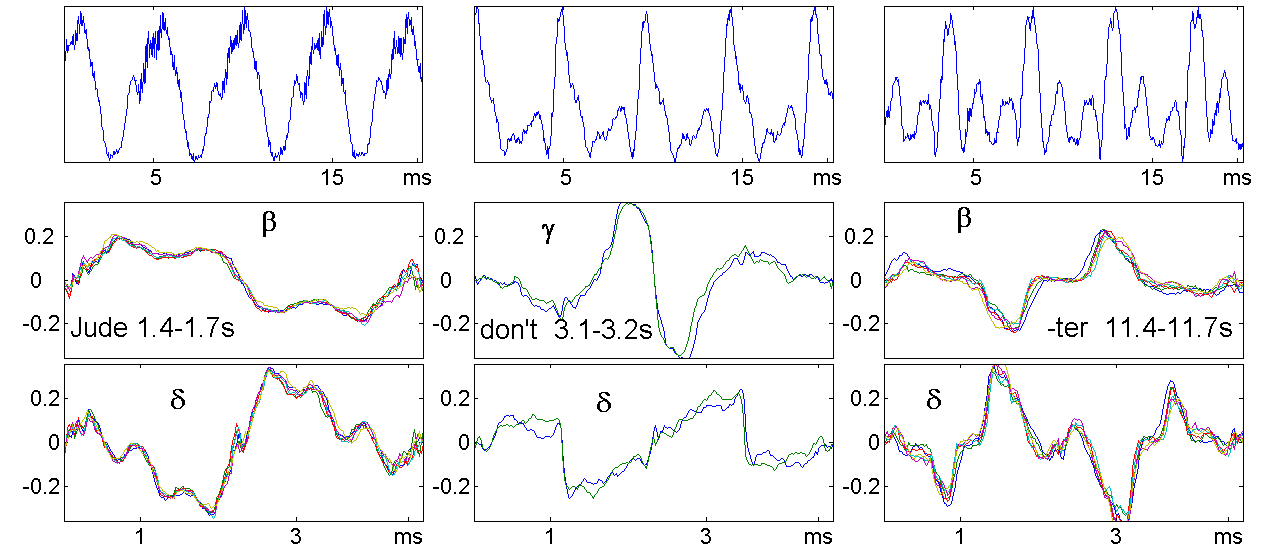}
\caption{Detail from Figure \ref{heyjude2}. The vowels of 'Jude', 'don't', and the second syllable in 'better' provide  stationary parts of the time series lasting for 0.3, 0.1, and 0.3 s. Only 20ms of each signal are shown in the top panel.  Ordinal autocorrelation functions were calculated for six resp. two disjoint windows of length 50ms and drawn for one pitch period, which equals 4.5ms for all three sounds. 
\vspace{3ex}}  \label{heyjude3}

\begin{minipage}{\textwidth}
The study of these patterns can improve methods of speech synthesis, speech analysis, and speaker authentification, but this is not an easy matter.  There is a huge variety of $\beta , \delta ,$ and $\gamma$ shapes which have to be collected, classified and understood. And there are open questions.  Is there an 'inverse transform' which reconstructs the time series from its ordinal autocorrelation functions?  What are the connections between power spectrum, cepstrum and ordinal functions? Can a speaker reliably reproduce the $\beta$-shape?
Can we hear the ordinal functions? If not, will the computer understand human speech better than humans themselves?
\end{minipage}
\end{figure}

\begin{figure}[ht]
{\bf 5. Weather, dust, and tides}\vspace{1ex}
\begin{minipage}{\textwidth}
We conclude with some everyday data.  The German Weather Service (\url{www.dwd.de}, Climate and Environment, Climate Data) provides hourly values of earth temperature, at depth 5cm, starting 1978. Can I learn something about my town?  Figure \ref{temp} shows the last 2000 values of the year 2013. In  autumn, there is still a daily rhythm, in winter this is rarely the case. We determine the persistence of the dataset with sliding windows of length 720, that is, one month. The effect of summer and winter can be seen, and we also see some irregularity: the data were first sampled every 3 years (which one can guess from the picture), then three times a day, and since 2001 every hour. For such plausibility checks even the short window is enough. However, there is no chance to see details in different months. The resolution of data is not sufficient, even for this short window which already causes statistical inaccuracy.
\vspace{3ex}
\end{minipage}
\includegraphics[width=.95\textwidth]{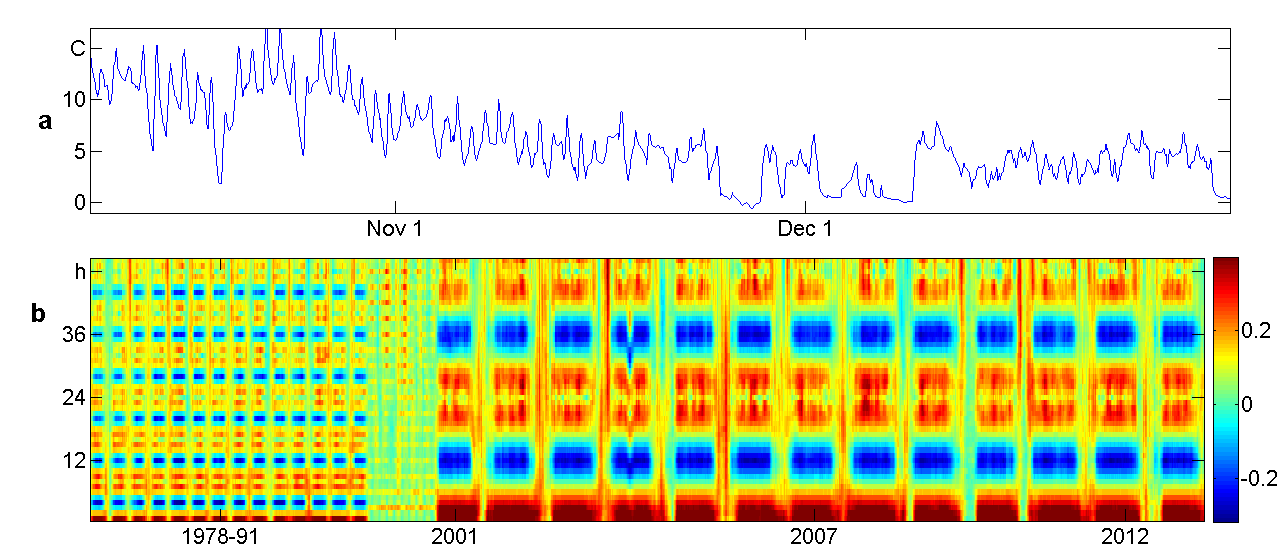}
\caption{Hourly temperature of earth in Greifswald, Germany. {\bf a.} The last 2000 values of 2013 show that the daily cycle is weak in winter. {\bf b.} Persistence for $d=1,...,50$h shows irregularities in the data.}  \label{temp}\vspace{3ex}
\begin{minipage}{\textwidth}
Hourly data of PM10 particulate measurements were already used in Figure \ref{corrfcts}. These are notoriously noisy data, even if we consider them on logarithmic scale, as seen in the upper part of Figure \ref{dustclust}. Particulates in the air are not uniformly distributed, they come in fractal clusters. Actually these data, together with the EEG, have the smallest $\Delta^2$ values in our studies.  In the dataset of San Bernardino from \cite{dust}  (station 3215 Trona-Athol, the other station 3500 shows different characteristics) the mean $\Delta^2$ is 0.003. For our window length 1200, there were 96\%  of small $\Delta^2$ values according to our bound.  Moreover, there were 13\% missing values.  

For screening the data, we omit all time points with missing values. This is not quite correct, but worth a trial, and the whole financial world works with disrupted time series. Persistence and up-down balance in Figure \ref{dustclust} show that in summer, but not in winter there is a daily rhythm in the data. It is similar as for temperatures above, but not obvious in the data. So the curves in Figure \ref{corrfcts}b are an effect of summer only. \vspace{3ex}
\end{minipage}
\end{figure}

\begin{figure}[h]
\includegraphics[width=.8\textwidth]{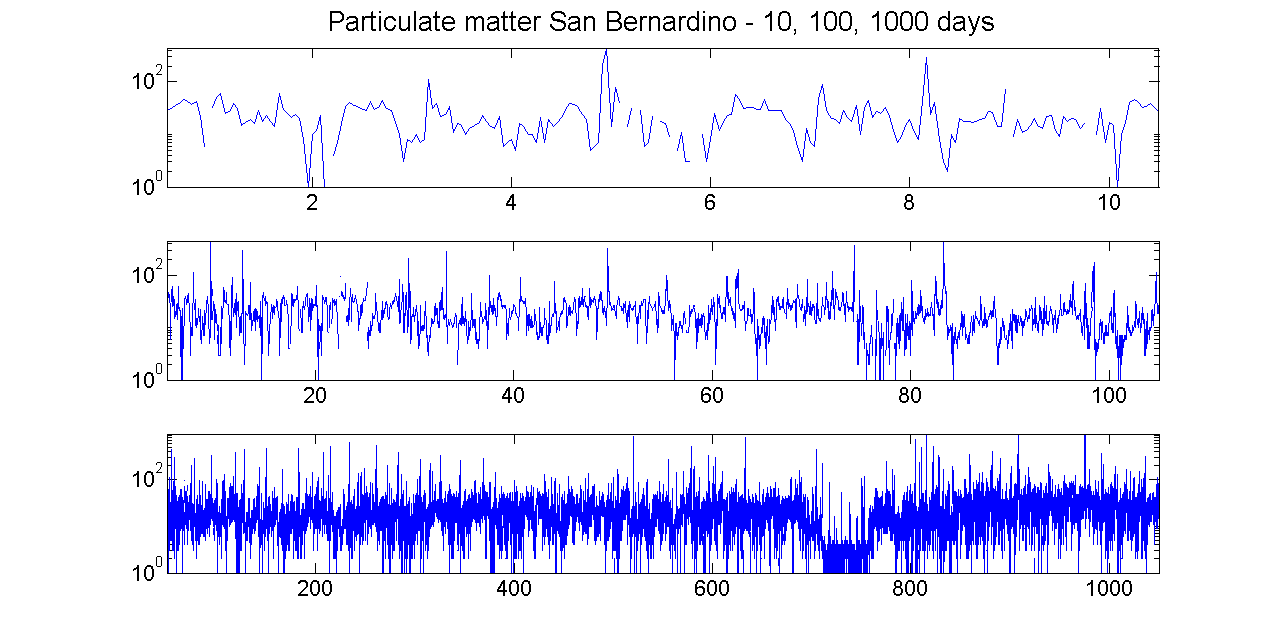}
\includegraphics[width=.99\textwidth]{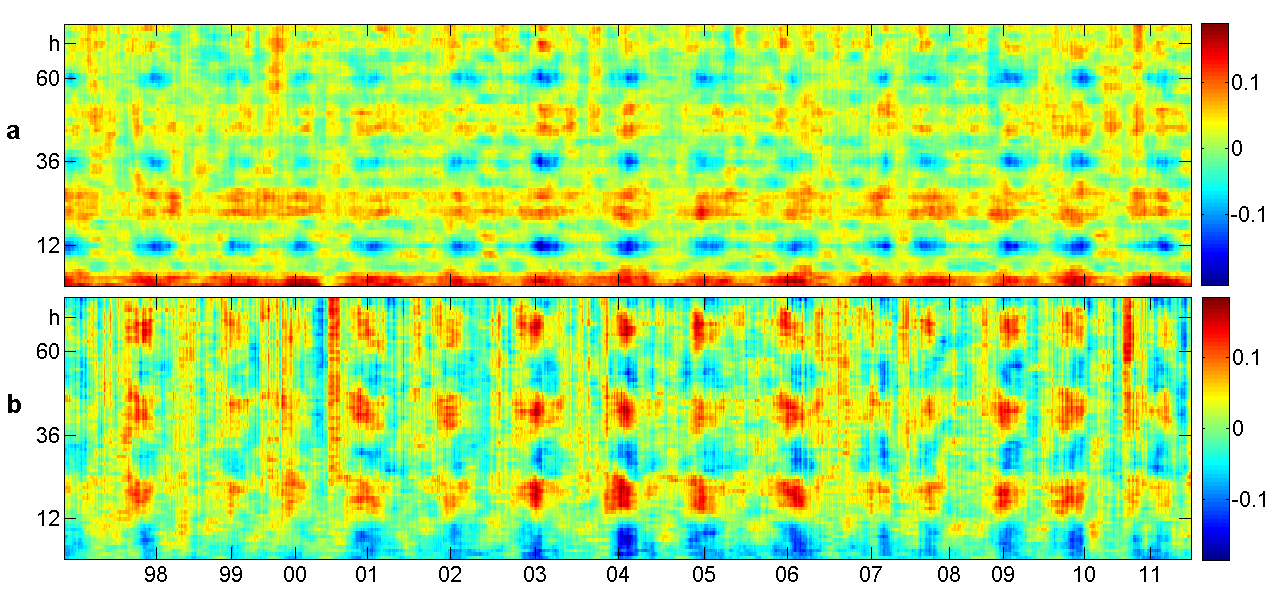}
\caption{Hourly PM10 measurements at San Bernardino, California \cite{dust}. The upper panel demonstrates the irregular character of the time series.  Ignoring missing data, {\bf a} persistence $\tau$ and {\bf b} up-down balance  $\beta$ were determined with a sliding window of length 1200. Years are marked at June 1. There is a daily rhythm in summer which is hardly visible in the data or in autocorrelation, cf. Figure \ref{corrfcts}. Hourly resolution does not allow to go further into detail. \vspace{3ex}}\label{dustclust}

\begin{minipage}{\textwidth}
Based on this impression we can now decide how to proceed the data further. It is not possible, however, to go into detail in Figure \ref{dustclust}. With a window of 50 days we can not see changes between weeks.  The sensor measurements are done every few seconds, however, and only the hourly averages are kept.  With data of finer resolution, say every 6 minutes, much more detail could be studied.

This is demonstrated in Figure \ref{tides} for tidal data from Anchorage, providing one more verification for our partition formula. Of course the tides form a very regular process, with the highest $\Delta^2$ of all our examples.  The possibility to work with a window of 5 days instead of several weeks, and the resulting details in the figure are due to the high resolution of the data, however. So this note should send a message to colleagues who design measuring equipment: Do not preselect data. Give users an option to study the wealth of all measured values and make their own choices. 
\vspace{3ex}
\end{minipage}
\end{figure}

\begin{figure}[ht]
\includegraphics[width=.8\textwidth]{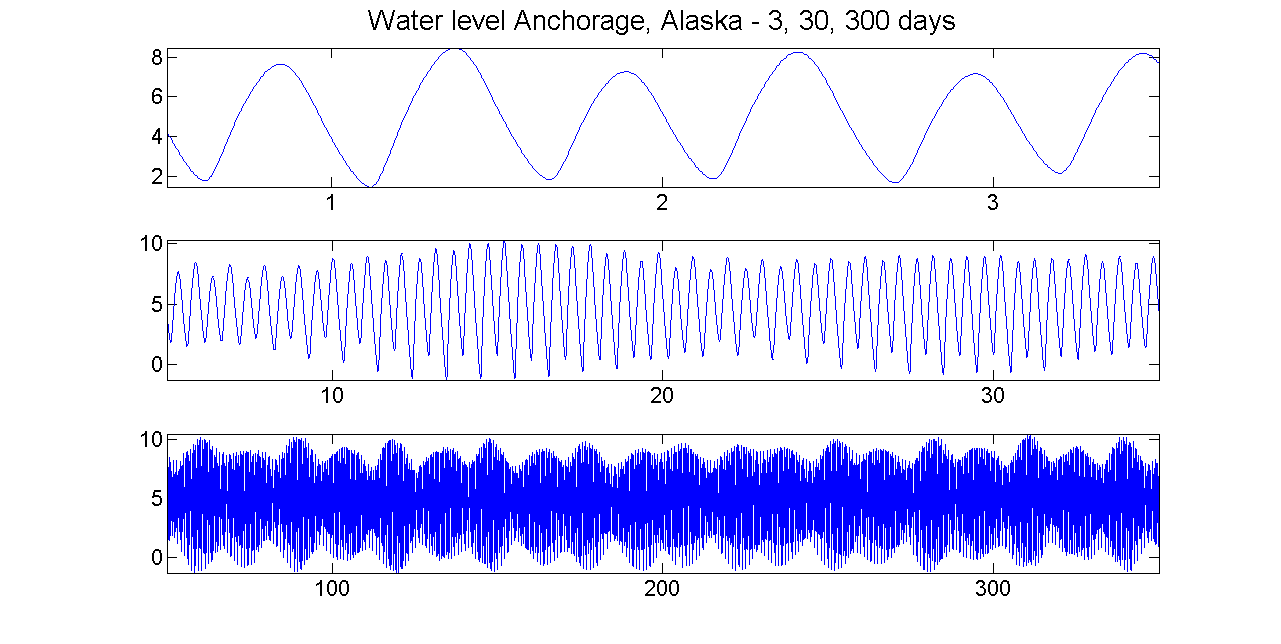}
\includegraphics[width=.99\textwidth]{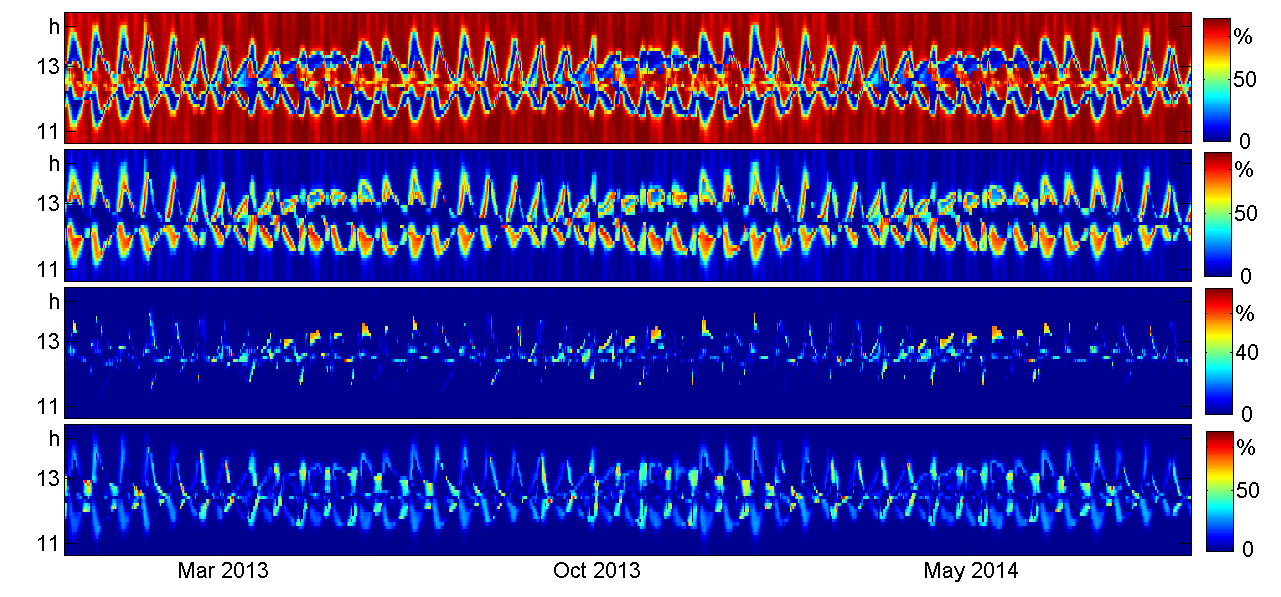}
\caption{We used tidal data to show that there are interesting $\beta$-shapes for different sites, by averaging over a whole month. The main actors in the tides' play are moon and sun, however, and the data have a multi-scale structure, shown in the upper panel. It seems interesting that ordinal functions make monthly and yearly cycles more transparent than the series itself, autocorrelation and spectrum. 
The partition of $\Delta^2$ for water levels from Anchorage was restricted here to the rather short time from Jamuary 2013 to September 2014, and to delays $d$ between 10.5 and 14.5 hours, around half a lunar day. A window of 1242 values, 5 lunar days, was shifted in steps of 12.4 hours. There are only 2\% of places with $\Delta^2<\frac{15}{n}.$ If they are excluded, the average value of  $\tilde{\tau}$ is 76\%, followed by  $\tilde{\beta}$ with 15\%,  $\tilde{\delta}$  with 6\%, and  $\tilde{\gamma}$ with 2\%. The average error is 0.54\% . }  \label{tides}\vspace{10ex}
\end{figure}

\begin{figure}[h]
{\bf 6. A fast laser experiment}\vspace{1ex}
\begin{minipage}{\textwidth}
The problem in this last section is rather simple: find the period of a very noisy rhythmic process with extremely short windows. We include this example because the data show what can be measured with current technology within parts of a millisecond.\vspace{3ex}
\end{minipage}

\includegraphics[width=.99\textwidth]{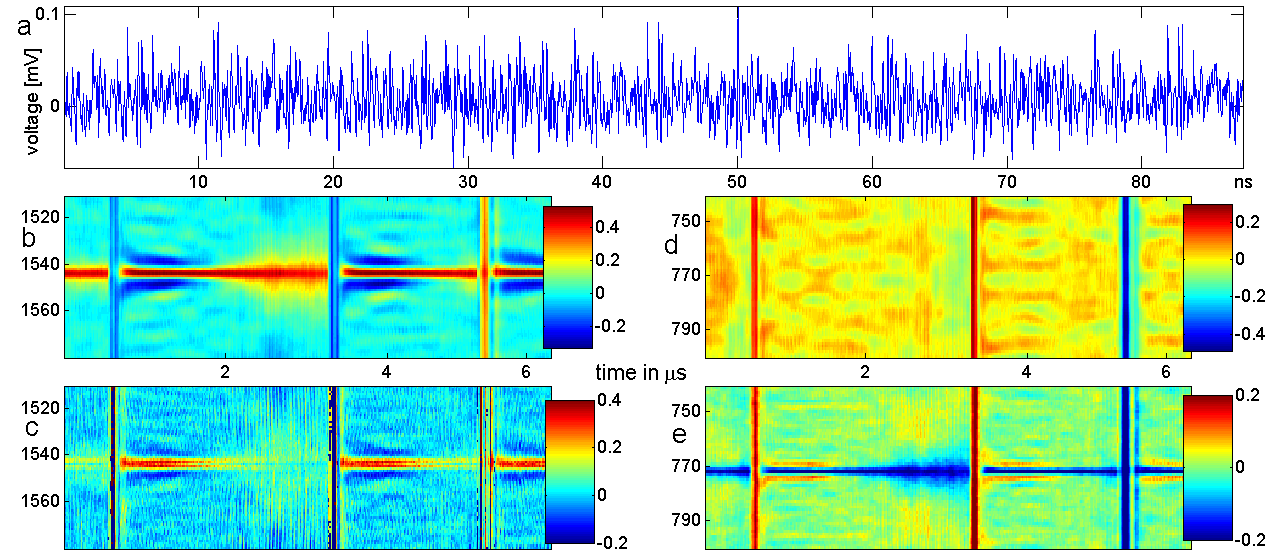}
\caption{Data for this figure come from the laser experiment of Sorriano et al.~\cite{sz}. A laser with wavelength 1542 nm and 11.7 mA threshold current was excited just above the threshold with 12 mA.  The sampling frequency was 25ps which means that 240000 values were measured during $6\mu s.$ The optical feedback was arranged so that the signal has a period of $L=1544.$ We try to find this period with a window of 2.3 periods, or 88 ns, just the length of the series shown in {\bf a}, where periodicity can hardly be seen. In {\bf b} and {\bf c}  we study autocorrelation and persistence around $d=L.$ The vertical interruptions show low-frequency disturbances, the laser does not run smoothly.  {\bf d} and {\bf e} show $\rho$ and $\tau$ around $d=\frac{L}{2}.$ While $\rho$ is a bit better at the full wavelength, $\tau$ detects the periodicity reliably at the half wavelength. }  \label{laser3500}
\end{figure}

\end{document}